\documentclass[aps,prd,amsmath,amssymb,superscriptaddress, preprintnumbers,preprint,nofootinbib,a4paper]{revtex4}
\pdfoutput=1

\usepackage[utf8x]{inputenc}
\usepackage{amsmath}
\usepackage{amsfonts}
\usepackage{amssymb}
\usepackage{graphicx,amsfonts,color,comment,amsmath,hyperref,float}
\usepackage{bm}% bold math
\usepackage{cancel}
\usepackage{enumitem}
\usepackage{slashed}

\usepackage{color}

\newcommand{\beq}{\begin{eqnarray}}
\newcommand{\eeq}{\end{eqnarray}}
\newcommand{\GeV}{\,\text{GeV}}
\newcommand{\TeV}{\,\text{TeV}}

\def\mg5{\textsc{MG5\_aMC\@NLO}}
\def\Pythia8{\textsc{Pythia8}}

\def\eq#1{{eq.~(\ref{#1})}}

\def\fig#1{{Fig.~(\ref{#1})}}
\def\sec#1{{Section ~\ref{#1}}}

\def\app#1{{Appendix ~\ref{#1}}}

\newcommand{\be}{\begin{equation}}
\newcommand{\ee}{\end{equation}}
\newcommand{\bea}{\begin{eqnarray}}
\newcommand{\eea}{\end{eqnarray}}
\newcommand{\nn}{\nonumber}

\begin{document}

\title{Top-quark Partial Compositeness beyond the effective field theory paradigm }

\author{Diogo Buarque Franzosi}
\affiliation{
 Department of Physics, \\ Chalmers University of Technology, \\ Fysikg\aa rden 1, 41296 G\"oteborg, Sweden}

\author{Alberto Tonero}
\affiliation{Ottawa-Carleton, Institute for Physics, Carleton University
1125 Colonel By Drive, Ottawa, ON, K1S 5B6, Canada} 
 
\begin{abstract}
In theories of Partial Compositeness the top quark is a mixture of a composite and an elementary state, and as a consequence its interactions with gauge bosons are expected to deviate from those of a point-like object.
At sufficiently large energies, such deviations cannot be parametrized by the leading effective field theory operators and form factors (\emph{i.e.} energy dependent interactions) must be introduced. 
In this work, we argue that such effects might appear at relatively low energies with interesting phenomenological consequences. 
In analogy to the proton electromagnetic interactions, we devise a simplified phenomenological model that parametrizes the top-quark interactions with gluons in terms of two form factors. We study the implications of these interactions in top-quark and heavy top-partner pair production at a hadron collider.
\end{abstract}
 
\maketitle

%%%%%%%%%%%%%%%%%%%%%%%

%\tableofcontents

%%%%%%%%%%%%%%%%%%%%
\section{Introduction}
%%%%%%%%%%%%%%%%%%%%%%%
The Standard Model (SM) of particle physics is a successful theory that describes with great  accuracy the vast majority of high energy data at our disposal. However, there are strong experimental evidences (dark matter, neutrino masses, baryon asymmetry) together with theoretical prejudices (hierarchy or naturalness problem) which indicate that the SM is an incomplete theory.

Physics beyond the SM is needed in order to explain these phenomena and to solve the naturalness problem. For what concerns the hierarchy problem, composite Higgs models are among the most promising theories. In the composite Higgs framework, a new confining dynamics is responsible for the Higgs boson compositeness and this has the merit of solving the hierarchy problem, since the Higgs mass scale is dynamically generated. Furthermore, the breaking of the electroweak (EW) symmetry also arises dynamically, in contrast to the SM where it is merely described by a ``wrong-sign'' mass term of the Higgs potential.

In these models, the generation of a sizeable top-quark mass is particularly challenging. The most promising ingredient to generate the correct fermion masses and the SM flavor structures is the concept of partial compositeness~\cite{Kaplan:1991dc}.
In models with partial compositeness, the top-quark is usually considered to be a mixture of one (or more) composite state $T'$ and an elementary state $t'$. Partial compositeness provides a compelling mechanism to give the correct mass to the top quark, as long as the composite operators that mix with the top have large anomalous dimensions.
%In the PC framework, the top-quark is a mixture of the elemenary fermion states $t'$ and a composite fermionic states $T'$ from the strongly interacting sector. 
Such construction has been studied in the context of 4 dimensional gauge theories with new hyperfermions charged under the hypercolor group~\cite{Ferretti:2016upr,Ferretti:2013kya,Ferretti:2014qta,BuarqueFranzosi:2019eee}.

Composite objects interact with gauge boson differently than elementary states. The best evidence of this fact can be seen in the interactions of nucleons with the electromagnetic field. Indeed, the most general form of the hadronic current $J^{\mu}_N$ for a spin $1/2$ nucleon with internal structure, satisfying relativistic invariance and current conservation is\footnote{The current is derived considering the complete on-shell nucleon line, this simplifies the Lorentz and Dirac structure of the interaction vertex.} \cite{Foldy:1952zz}:
\begin{equation}\label{emff}
J^{\mu}_N=e\bar{N}(p')\left[\gamma^{\mu}F^N_{1}(Q^{2})+\frac{i\sigma^{\mu\nu}q_{\nu}}{2M_N}F^N_{2}(Q^{2})\right]N(p)
\end{equation}
where $M_N$ is the nucleon mass, $Q^{2}=-q^{2}>0$ and $q$ is the photon momentum $q=p'-p$. 
The  dimensionless functions $F^N_{1}$ and $F^N_{2}$ are the so called Dirac and Pauli form factors.

In complete analogy to the nucleon case, in this work we parametrize the interactions of the heavy composite top partner with gluons in terms of two form factors and study the phenomenological consequences. 

The paper is organized as follows. In \sec{sec:CD} we describe the composite dynamics and the class of models we are going to consider. We present a simplified model of partial compositeness that features all necessary ingredients for our study.
In \sec{sec:onegluonff} we introduce two form factors that parametrize the interactions of the heavy top partner $T'$ to a single gluon field and we derive the corresponding top quark form factors induced by the partial compositeness mechanism. We provide a phenomenological parametrization of these form factors in terms of some parameters 
and discuss the interplay of the relevant scales in describing this modified interactions, taking hadron physics as guidance. We finally discuss the implications of these interactions for $q\bar{q}$-initiated top partner and top-quark pair production.
In \sec{sec:gg} we expand the prescription to the $gg$-initiated process $gg\to t\bar{t}$, using as guidance the modeling of proton-antiproton production in photon-photon scattering. We  conclude in \sec{sec:conclusion}.

%%%%%%%%%%%%%%%%%%%%%%
\section{Composite dynamics}
\label{sec:CD}
%%%%%%%%%%%%%%%%%%
As the name indicates, \emph{partial compositeness} refers to a class of models in which a composite state $T'$ mixes with an elementary top quark state $t'$ generating its mass. 
The simplest and most straightforward UV completion of this idea is provided by four-dimensional purely fermionic gauge theories, which accommodates a Higgs particle as fermionic bound state of pseudo Nambu-Goldstone Boson nature. 
In these type of models the SM is extended by a new hypercolor gauge group $G_{HC}$ with new EW charged hyperfermions $\psi$ that condensate and 
spontaneously break the global symmetry of the theory, including the EW group. Then, via the vacuum misalignment mechanism~\cite{Dugan:1984hq} the EW scale  $v=246\GeV$ is naturally and dynamically generated 
\begin{equation}
v=f\sin\theta
\label{eq:misangle}
\end{equation}
with $f$ the decay constant of the Nambu-Goldstone Boson of the symmetry breaking generated by the vacuum condensate and $\theta$ the misalignment angle. 

Top partner candidates are usually composed of three hyper-fermions charged under $G_{HC}$ as well as EW and QCD. They might belong to the same representation of $G_{HC}$~\cite{Vecchi:2015fma} or to two different representations~\cite{Ferretti:2013kya,Ferretti:2014qta,Ferretti:2016upr}.
Extensions of this framework includes hypercolor charged scalars in which case top partners would be composed of one hyperscalar and one hyperfermion~\cite{Sannino:2016sfx}.

Regardless of the specific structure of the $T'$ state, its gauge interactions will be deformed with respect to a point-like particle. Here we adopt a model independent view and consider as benchmark a simplified model of partial compositeness in which all information about its constituents is embedded in the gauge interaction form factors.
This simplified model contains all necessary ingredients for our phenomenological discussion and will be presented in the following sections.

%%%%%%%%%%%%%%%%%%%%%%
\subsection{Simplified partial compositeness model}
\label{sec:PC}
%%%%%%%%%%%%%%%%%%
In order to show how the PC mechanism works for the case of the top-quark we consider a simplified model that features one single vector-like composite top partner $T'$ transforming as a  $ {\bf (3,1)_{2/3}}$ under SU(3)$\times$ SU(2)$_L\times$U(1)$_Y$. Furthermore, let $Q'_L=(t'_L, b'_L)\sim {\bf (3,2)_{1/6}}$  and $t_R' \sim {\bf (3,1)_{2/3} }$ be the SM elementary third generation left quark doublet and the right handed top quark, respectively. We consider the following mixing terms
\begin{equation}\label{PClagr}
{\cal L} \supset -M \bar{T'}_L T'_R - y \bar{Q'}_L \widetilde{H} T'_R - \lambda f \bar{T'}_L t'_R + \text{h.c.}\,
\end{equation}
that are built with up to one insertion of the Higgs doublet $H$ ($\widetilde{H}=i\sigma_2 H^*$). 
Here $M$ is the mass of the heavy top partner $T'$, $y$ and $\lambda$ are parameters that can be computed in terms of the four-fermion interactions in the UV theory, 
and $f$ and $\theta$ are defined in \eq{eq:misangle}. Electroweak precision data and Higgs coupling measurements require a scale $f\gtrsim 1\TeV$~\cite{Barbieri:2012tu,Grojean:2013qca,Arbey:2015exa}, although it has been argued that the contribution of other  composite states might alleviate that bound to $f\gtrsim 600\GeV$~\cite{BuarqueFranzosi:2018eaj}.

After EW symmetry breaking $\langle H \rangle_0 = (0\,,\, \upsilon/\sqrt{2})$ we can identify the following mass mixing terms 
\vspace{0.1cm}
\begin{equation}\label{Lmass}
-{\cal L}_{\rm mass} = (\bar{t'}_L \quad \bar{T'}_L)\binom{0\quad \frac{y \upsilon}{\sqrt{2}} }{\lambda f\quad M}\binom{t'_R}{T'_R} + \text{h.c.}
\end{equation}
Let $t$ and $T$ be the mass eigenstates such that
\be \label{Ttdiagon}
\binom{t'_R}{T'_R}=\binom{-c_R\quad s_R}{s_R\quad c_R}\binom{t_R}{T_R}\qquad {\rm and}\qquad
\binom{t'_L}{T'_L}=\binom{c_L\quad s_L}{-s_L\quad c_L}\binom{t_L}{T_L}
\vspace{0.3cm}
\ee
where $c_{R,L}=\cos\theta_{R,L}$ and $s_{R,L}=\sin\theta_{R,L}$. 
To achieve a diagonal matrix we get
\begin{equation}
\tan(2\theta_L)=\frac{\sqrt{2}Myv}{M^2-\frac{y^2v^2}{2}+\lambda^2f^2},\quad\quad \tan(2\theta_R)=\frac{2M\lambda f}{M^2+\frac{y^2v^2}{2}-\lambda^2f^2}\,.
\end{equation}
At leading order in 
$\upsilon/f$ we have that
\be 
c_R\simeq\frac{M}{\sqrt{\lambda^2f^2+M^2}},\quad s_R\simeq\frac{\lambda f}{\sqrt{\lambda^2f^2+M^2}}, 
\quad c_L\simeq 1,\quad s_L\simeq \frac{y\upsilon}{\sqrt{2}}\frac{M}{\lambda^2f^2+M^2}
\label{eq:mixangle}
\ee
and the masses of $t$ and $T$ are given by
\be \label{masses}
m_t\simeq\frac{y\upsilon}{\sqrt{2}}\frac{\lambda f}{\sqrt{\lambda^2f^2+M^2}}\qquad {\rm and}\qquad
m_T\simeq \sqrt{\lambda^2f^2+M^2}\,.
\ee
The number of free parameter in eq.~\eqref{PClagr} are four but thanks to the top mass relation in eq.~\eqref{masses} we can reduce them down to three that we take to be $M$, $\lambda$ and $f$. 
By inverting the relation for $m_t$ we get
\be 
y=\frac{\sqrt{2}m_t}{\upsilon}\sqrt{1+\frac{M^2}{\lambda^2f^2-m_t^2}}\simeq \sqrt{1+\frac{M^2}{\lambda^2f^2}}\,.
\ee

%%%%%%%%%%%%%%%%%%%%%%%%%%%
\section{One-gluon phenomenological form factors}
\label{sec:onegluonff}
%%%%%%%%%%%%%%%%%%%%%%%%
The heavy top partner $T'$ we introduced in the previous section is considered to be a fully composite object (like a nucleon), made of hyperquarks, and therefore it interacts with gluons differently than a point-like particle. In the absence of mixing we can write, in analogy to eq. \eqref{emff},  the current that parametrizes the interaction of an on-shell $T'$ with a single gluon $G_\mu^a$ as follows~\footnote{see \app{app:genericFF} for the discussion about the derivation of the most general form of the current.}
\be \label{TprimeJ}
(J_{T'})^{\mu,a}=g_{s}\bar{T'} T^{a}\left[\gamma^{\mu}F_{1}(q^{2})+\frac{i\sigma^{\mu\nu}q_{\nu}}{2M}F_{2}(q^{2})\right]T'
\ee
where $F_1(q^2)$ and $F_2(q^2)$ are the Dirac and Pauli chromo form factors of $T'$, $q^2$ is the virtuality of the  gluon and $T^a$ are the SU(3) generators. Notice that the vector-like nature of QCD is respected by the form of the current in eq.~\eqref{TprimeJ}. This current resembles very closely the EM hadronic current of nucleons (in our case the role of the photon is taken by the gluon and the role of the nucleon is taken by the heavy top) and can be derived from a gauge invariant lagrangian as shown in Appendix \ref{app:gaugeinvff}. Furthermore, the Dirac form factor can be written as 
\be \label{F1express}
F_{1}(q^{2})=1+\frac{q^2}{M^2}f_1(q^2)\,,
\ee
in order to ensure the correct gauge charge normalization.
At zero momentum transfer the composite state $T'$ behaves as a coherent sum of its constituents without structure and this fixes the value of the form factors at $q\to 0$,
\be \label{Fzero}
F_{1}(0)=1\qquad{\rm and}\qquad F_{2}(0)=\kappa_g\,,
\ee
where $\kappa_g$ is the anomalous chromomagnetic dipole moment of $T'$. Thanks to the PC mechanism described in Section \ref{sec:PC}, the light mass eigenstate $t$ (which is a mixture of the fully composite top partner $T'$ with the fundamental top $t'$) will present as well non standard interactions with gluon parametrized by the following current 
\be \label{tJ}
(J_t)^{\mu,a}=g_{s}\bar t T^{a}\left[\gamma^{\mu}F_1^{tg}(q^2)+\frac{i\sigma^{\mu\nu}q_{\nu}}{2M}F_{2}^{tg}(q^{2})\right]t
\ee
where the top-quark form factors are now given by
\be \label{topff1}
F_1^{tg}(q^2)=1 +(s_L^2 P_L + s_R^2 P_R)\frac{q^2}{M^2} f_{1}(q^{2})
\ee
and
\be \label{topff2}
F_{2}^{tg}(q^{2})=-s_Ls_RF_{2}(q^{2})\,.
\ee
In eq.~\eqref{topff1}-\eqref{topff2} $P_L=\frac{1}{2}(1-\gamma_5)$ and $P_R=\frac{1}{2}(1+\gamma_5)$ are the left- and right-handed projectors while $s_L$ and $s_R$ are the sine of the left and right mixing angle introduced in eq.~\eqref{Ttdiagon}. 
Similar form factors are generated for the mass eigenstate top partner $T$. Notice that, due the PC mechanism, non vector-like interaction terms appear in the Dirac form factor in eq.~\eqref{topff1}. The vector-like nature of the strong interactions is recovered  at zero momentum transfer, where the gauge symmetry fixes the form of the interaction as well as in the special limit $s_L=s_R$.
Finally, besides gluon interactions with $t\bar t$ and $T\bar T$, we expect also gluon interactions with $t\bar T$ and $T\bar t$ to be present. We simply point out that these are generated at higher momenta and represent a new form of having single top-partner production via QCD interactions that should be further investigated.

%%%%%%%%%%%
\subsection{Phenomenological parametrization of the form factors}
\label{sec:ff}
%%%%%%%%
Modeling nucleon EM form factors has been a long-standing activity that we are going to borrow in order to describe the compositeness of the heavy top partner. 
%%%%%%%%%%%
\subsubsection{The proton case}
%%%%%%%%
Before discussing the top quark, it is useful to reconsider first the nucleon case 
and introduce the well known \emph{Sachs} electric and magnetic form factors $G_E$ and $G_M$ which  are defined as follows
\begin{eqnarray}
G_{E}^N(q^{2}) & = & F_{1}^N(q^{2})+\frac{q^{2}}{4M_p^{2}}F_{2}^N(q^{2}) \,,\\
G_{M}^N(q^{2}) & = & F_{1}^N(q^{2})+F_{2}^N(q^{2})\,.
\label{eq:sachs}
\end{eqnarray}
Data from several proton scattering experiments are well fitted by the so-called dipole approximation~\cite{Iachello:1972nu}
\begin{equation}
G_E^p(q^2)= \mu^{-1}G_M^p(q^2) = G_D^p(q^2)=\left(1-q^2/m_D^2 \right)^{-2}\,
\end{equation}
where $m_D^2=0.71\GeV^2$ and $\mu=1+\kappa\approx 2.8$ is the magnetic moment of the proton, $p$. This approximation describes very well the data in the space-like region $(q^2<0)$, however in the time-like region $(q^2>0)$ an absorptive factor as well as a bunch of resonances need to be introduced to fit properly the data (for instance of proton pair production in $e^+e^-$ collisions)~\cite{Iachello:2004aq,Bijker:2004yu}. If one is mostly interested in the low energy behavior of the form factors, it is possible to neglect the effects of resonances, being sufficient to introduce an absorptive phase $\theta$ as follows
\begin{equation}
G_D^p(q^2)=\left(1-\frac{q^2}{m_D^2}e^{i \theta\Theta(q^2)} \right)^{-2}
\end{equation}
with $\Theta(q^2)=1$ if $q^2>0$ and 0 otherwise. This approximation is valid below threshold, for $q^2<m_\rho^2$, where
$m_\rho$ is the mass of the lightest QCD resonance (the rho meson). The expansion of the form factors around $q^2=0$ defines the radius of the nucleon 
\be \label{expansionsachs}
G_{E,M}^p(q^{2})  = G_{E,M}^p(0)\left( 1+\frac{q^2}{6}\langle r^2\rangle_{E,M} +\ldots\right)
\ee
and its compositeness scale $\Lambda_c^p$ such that
\be 
\frac{(\Lambda_c^{p})^2}{6}\langle r^2\rangle\sim 1\,.
\ee
In terms of the dipole approximation we have that $\Lambda_c^p\approx 560$ MeV. This scale is related to the \emph{compositeness} of the object, \emph{i.e.} it represents the scale above which the proton constituents start to be seen as individual objects. Notice that it is typically smaller than the chiral symmetry breaking scale $\Lambda_{\chi}=4\pi f_\pi\simeq 1$ GeV as pointed out by Manohar and Georgi in their non-relativisitc quark model (NRQM)~\cite{Manohar:1983md}. 

Let us conclude the discussion about the proton compositeness by presenting the explicit form of $F_1^p$ and $F_2^p$ in the dipole approximation
\begin{eqnarray}
F_1^p(q^2)&=&G_D^p(q^2)\left(1+\frac{\kappa}{1-4M_p^2/q^2}\right) \nonumber\\
F_2^p(q^2)&=&G_D^p(q^2)\left(\frac{\kappa}{1-q^2/(4M_p^2)} \right)\,.
\label{eq:DAF}
\end{eqnarray}
Notice that, although the anomalous magnetic moment is a typical property of a composite object, it is also the property of its coherent sum and thus it does not define the compositeness scale of the object, as the radius does. It is also important to notice that the definitions in \eq{eq:sachs} become degenerate at $q^2=4M_P^2$ and this leads to the appearance of a pole in the form factor expressions in \eq{eq:DAF}. These poles however are not physical, they are expected to be removed from other (non-dipole) contributions and they do not spoil the high energy behavior.  
%%%%%%%%%%%
\subsubsection{The top-quark case}
%%%%%%%%
In complete analogy to the proton case, see eq.~\eqref{eq:DAF}, we parametrize the gluonic form factors of the heavy $T'$  using the dipole approximation as follows
\bea \label{F12dapprox}
F_1(q^2)&=&\left(1-\frac{q^2}{M_D^2}e^{i \eta\Theta(q^2)} \right)^{-2}\left(1+\frac{\kappa_g}{1-4M^2/q^2}\right) \nn\\
F_2(q^2)&=&\left(1-\frac{q^2}{M_D^2}e^{i \eta\Theta(q^2)} \right)^{-2}\left(\frac{\kappa_g}{1-q^2/(4M^2)} \right)\,.
\eea
We note that these forms might be different than the proton case depending on the specific structure of the composite objects and the charges of its constituents. Nevertheless, we expect the asymptotic behavior $F_1(q^2)\to q^{-4}$ and $F_2(q^2)\to q^{-6}$ at large $q$ to be the same as in the proton case~\cite{Brodsky:1973kr,Lepage:1980fj} and this is reproduced by the dipole approximation. We therefore see this modeling well motivated.

These form factors depend on two mass scales $M$ and $M_D$ and two dimensionless parameters $\kappa_g$ and $\eta$. It is useful to introduce an additional scale $M_\rho$, associated to the mass of the lightest resonance of the new composite sector, above which the form factors (in the time-like region) are no longer described by the dipole approximation. The composite state $\rho$ is typically a hypermeson, composed of two hyperquarks of the strong sector.
The scale $M$ is the mass of the composite top partner and is typically larger than $M_\rho$ and is smaller than the chiral symmetry breaking scale
\be 
M_\rho<M< 4\pi f\,.
\ee
This is true for QCD and other theories whose spetra have been measured in the lattice~\cite{Ayyar:2018zuk}.
$\kappa_g\approx \mathcal{O}(1)$ is the coherent magnetic structure of the $T$ constituents, $\eta$ is the absorptive phase and $M_D$ is a mass parameter that gives the typical form factor scale.

\begin{figure}
\includegraphics[width=0.8\textwidth]{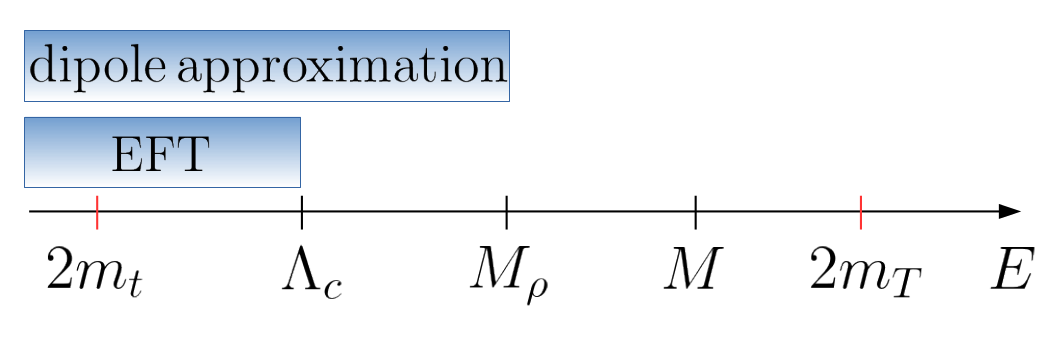}
\caption{Relevant scales and thresholds for the simplified partially composite top quark scenario discussed in the text.}
\label{fig:scales}
\end{figure}

The expansion of the form factors around $q^2=0$ defines the scale of compositeness $\Lambda_{c,i}$ 
\begin{equation}
F_i(q^2)\approx F_i(0)\left(1+\frac{q^2}{\Lambda_{c,i}^2}+\cdots\right)\,.
\end{equation}
The values of the form factors at zero momentum $F_i(0)$ are given in eq.~\eqref{Fzero} and
\be 
\Lambda_{c,1}^{-2}=\frac{2}{M_D^2}-\frac{\kappa_g}{4M^2}\,, \qquad\qquad\qquad \qquad
\Lambda_{c,2}^{-2}=\frac{2}{M_D^2}+\frac{1}{4M^2}.
\label{eq:lambdac}
\ee 
 Furthermore, by comparing the expansion of $F_1$ with eq.~\eqref{F1express} we have that
\be 
f_1(0)=\frac{M^2}{\Lambda_{c,1}^{2}}\,.
\ee
In this work we are interested in the case where compositeness effects are felt before the appearance of resonances, in complete analogy to QCD, namely
\be
\Lambda_c \lesssim M_\rho\,.
\ee
From here on we will use $\Lambda_c$ to collectively denote either $\Lambda_{c,1}$ or $\Lambda_{c,2}$, since in our scenario they are typically numerically close. As the typical energy scale of a certain physical process goes below $\Lambda_c$, only the coherent properties of the composite state are observable, which include its total (color) charge and its (chromo)magnetic moment. All these mass scales are shown in Fig.~\ref{fig:scales} together with the threshold production scales of two light and two heavy top quarks. Thanks to PC, the pair production threshold of partially composite top quarks tuns out to be smaller than the compositeness scale, namlely $2m_t<\Lambda_c$. This is different than the proton case where the proton is fully composite and $\Lambda_c^p<2M_p$.

%%%%%%%%%%%5
\subsection{Effective low energy expansion}
%%%%%%%%%%%%%%%
Combining the form factor expressions based on the dipole approximation given in eq.~\eqref{F12dapprox} with the form of the top quark interaction current resulting from the mixing with the heavy top partner given in \eq{tJ}, we have all the ingredients to compute physical processes involving on-shell top quarks. Before that, let us rephrase the results of the previous section in the language of effective field theories. 

The explicit form of the interaction current in \eq{tJ} can be derived from the following gauge invariant higher derivative  effective lagrangian (see \app{app:gaugeinvff})
\bea \label{ltopeff}
{\cal L}&=&\bar{t}i\gamma^{\mu}D_{\mu}t+\frac{g_{s}}{M^2}\bar{t}\gamma^{\mu}T_{a}(s_L^2 P_L + s_R^2 P_R)tf_{1}(-D^{2})D^{\nu}G_{\mu\nu}^{a}
-\frac{g_{s}}{4M}s_Ls_R\bar{t}\sigma^{\mu\nu}T_{a}tF_{2}(-D^{2})G_{\mu\nu}^{a}\nn \,.\\
\eea
Let us focus, for instance, on the dipole operator and consider the low energy EFT limit obtained by expanding the form factor $F_2$ around $-D^2=0$, the leading terms are
\be \label{eftops}
 -\frac{g_{s}\kappa_g}{4M}s_Ls_R\bar{t}\sigma^{\mu\nu}T_{a}tG_{\mu\nu}^{a}-\frac{g_{s}\kappa_g}{4M\Lambda_{c,2}^2}s_Ls_R\bar{t}\sigma^{\mu\nu}T_{a}t(-D^2G_{\mu\nu}^{a})+\ldots \,
\ee
This expansion breaks down at scales $E\sim \Lambda_{c,2}$ where the EFT description becomes no longer valid and the full expression of $F_2$ has to be used. As already said, thanks to the PC mechanism, the mass of the top quark turns out to be smaller than the compositeness scale, namely $m_t < \Lambda_{c,2}$, and therefore the process  $q \bar q \to g^* \to t\bar t$ can probe the form factors at low energies. At the same time, if $\Lambda_c$ is sufficiently low, collider experiments could access intermediate energies $\Lambda_c <  E < M_\rho$  where the top compositeness can be felt before the appearance of resonances (see Fig.~\ref{fig:scales}). We will return on this case in the next section where we will discuss specific benchmark scenarios. 

An extra subtlelty of the partially composite top model is related to the mixing angles that appear in the form factors. They enter only linearly in the EFT operators derived from the low energy expansion and this generates an extra enhancement of the higher dimensional operators with respect to the leading term. For instance the chromomagnetic dipole operator in the l.h.s. of eq.~\eqref{eftops} is suppressed by $\frac{s_L s_R\kappa_g}{ 4M}$ and the next-to-leading dimension-seven operator in the r.h.s. of eq.~\eqref{eftops} is suppressed only by $\frac{s_L s_R\kappa_g}{4 M\Lambda_{c,2}^2}$, instead of the naive EFT expectation $\frac{s_L^3 s_R^3 \kappa_g^{3} }{16M^3}$.

The novelty of our approach is to consider the full energy dependence of the form factors, which can be important in intermediate energy scales where $\Lambda_c <  E < M_\rho$ (see Fig.~\ref{fig:scales}). The assessement of the top structure at low energy via an EFT parametrization has been the only method applied so far~\cite{Englert:2012by,Englert:2014oea,Fabbrichesi:2013bca,Bernreuther:2013aga,Durieux:2018ekg}.  
The existing bounds on the EFT coefficients could be conservative if a proper analysis on the shape of the distributions which account for such energy dependent effects are not considered, and they are indeed weak for a typical motivated partial composite scenario.
We can obtain the leading dimension operators by taking the form factors at \eq{ltopeff} at zero momentum, and further use the dipole approximation to get $f_1(0)=M^2/\Lambda_{c,1}^2$ and $F_2(0)=\kappa_g$.
The marginalized bounds on the operators coefficients (considered separately) are~\cite{Franzosi:2015osa}\footnote{We identify $d_V=-s_L s_R \frac{m_t}{4M}\kappa_g$.}
\begin{equation}
-0.0099<-s_L s_R \frac{m_t}{4M}\kappa_g < 0.0123
\end{equation}
and\footnote{We use the result $-0.74\TeV^{-2}<\frac{C_1}{\Lambda^2}<0.71\TeV^{-2}$ and identify $\frac{C_1}{\Lambda^2}=g_s(s_L^2 +s_R^2)/2\Lambda_{c,1}^2$.}~\cite{Barducci:2017ddn}
\begin{equation}
-0.018< \frac{(s_L^2 + s_R^2)}{2} m_t^2\left(\frac{2}{M_D^2}-\frac{\kappa_g}{4M^2}\right) < 0.017\,.
\end{equation}

%%%%%%%%
\subsection{Probing the form factors in $q \bar q \to T\bar T$}
\begin{figure}
\includegraphics[width=0.49\textwidth]{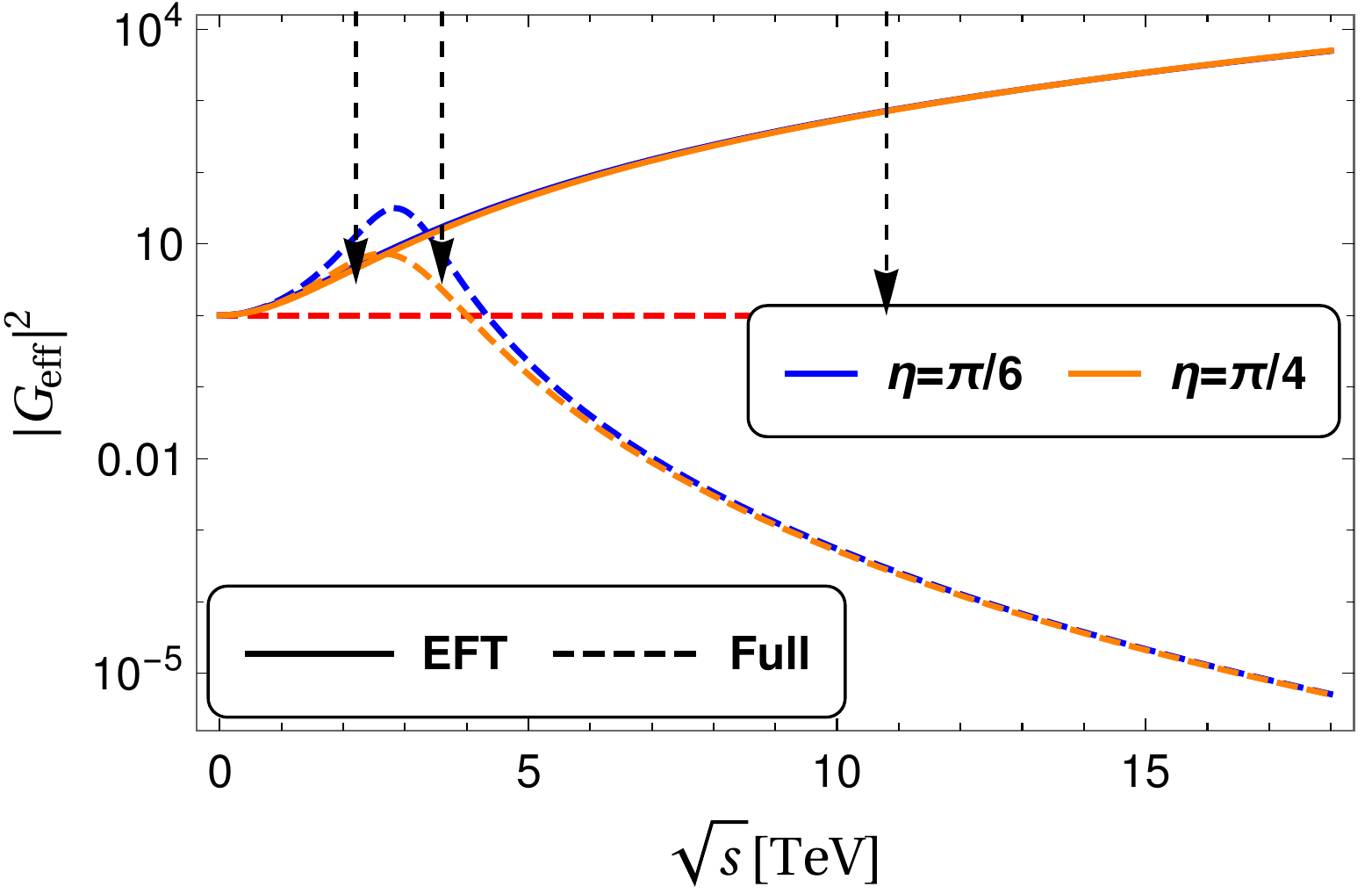}
\includegraphics[width=0.49\textwidth]{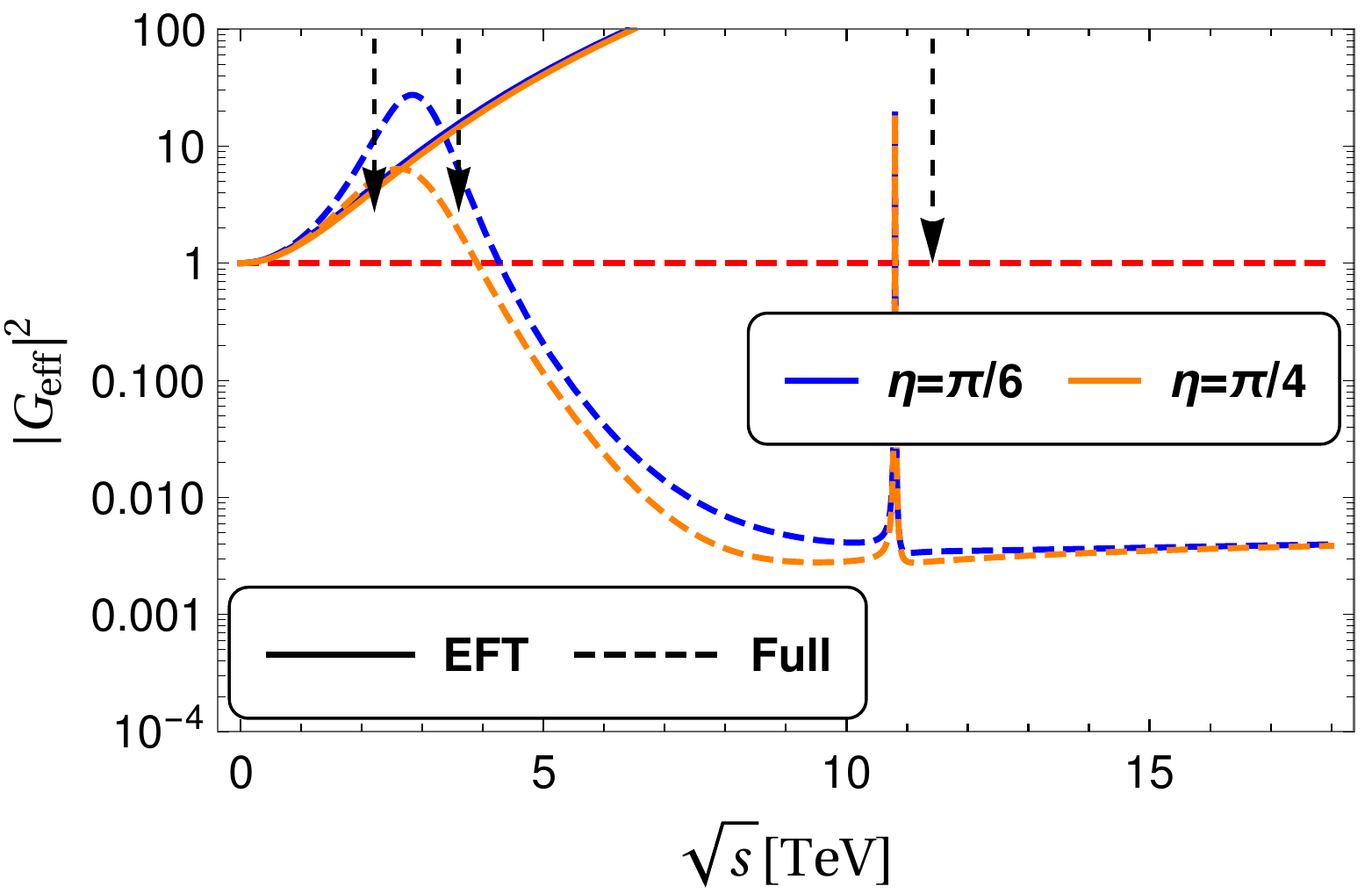}
\caption{Energy dependence of $|G_{eff,T}|^2$ computed for pair production of fully composite (left plot) and partially composite (right plot) heavy top states. In this case we fix $f=0.6\TeV$, $M=9 f$, $M_D=5f$, $\kappa_g=2$ and $\eta=\pi/6$.  The  arrows from left to right correspond to $\Lambda_{c}$, $m_\rho=6f$ and $2m_T$. }
\label{fig:Geff-pure}
\end{figure}

%%%%%%%%
Let us consider first the production of heavy top partners in the process $q \bar{q}\to g^*\to T\bar{T}$, which is a process that probes the top-quark form factors in the time-like region at high energies $E \gtrsim 2m_T \gg \Lambda_{c,i}$. The tree level cross section can be computed using the current in eq.~\eqref{TprimeJ} and taking into account the effect of mixing, it is given by
\be
\sigma_{q\bar{q}\to T\,\bar{T}}=\frac{8\alpha_{s}^{2}\pi}{27s}\sqrt{1-\frac{4 m_{T}^{2}}{s}}\left(1+\frac{2m_{T}^{2}}{s}\right)|G_{{\rm eff},T}(s)|^2
\label{eq:qqTT}
\ee
where 
\bea
|G_{{\rm eff},T}(s)|^2&=&1+\left( 1+\frac{2m_T^2}{s}\right)^{-1}\Bigg\{%\left( 1+\frac{2m_t^2}{s}\right)+
\frac{2m_T^2+s}{M^2}(c_R^2+c_L^2){\rm Re}f_1(s)+\frac{3m_T}{M}c_Rc_L{\rm Re}F_2(s)\nn\\
&&+
\frac{s^2}{2M^4}\left[(c_R^4+c_L^4)\left(1-\frac{m_T^2}{s}\right)+6c_R^2c_L^2\frac{m_T^2}{s}\right]|f_1(s)|^2+\frac{8m_T^2+s}{8M^2}c_R^2c_L^2|F_2(s)|^2\nn\\
&&+\frac{3m_T s}{2M^3}c_Rc_L(c_R^2+c_L^2){\rm Re}f_1(s)F^*_2(s)\Bigg\}\,.
\label{eq:GeffTT}
\eea
For illustrative purposes, in \fig{fig:Geff-pure} we show the energy dependence  of $|G_{{\rm eff},T}|^2$ assuming the dipole approximation to be valid up to energies $E \gtrsim 2m_T$. The left plot of \fig{fig:Geff-pure} shows the behavior of $|G_{{\rm eff},T}|^2$ (dashed blue curve) for a purely composite state ($\lambda=0$) for which $c_L=c_R=1$. In this case we fix  $f=0.6\TeV$, $m_T=M=9 f$, $M_D=5f$ and $\kappa_g=2$. We consider two values for the phase $\eta=\pi/4$ and $\eta=\pi/6$.
The right plot  of \fig{fig:Geff-pure} shows the behavior of $|G_{{\rm eff},T}|^2$ (dashed blue curve) where we have used instead $\lambda=3$ and derived the mixing angles and $m_T$ according to \eq{eq:mixangle} and \eq{masses}.
The  value $M=9 f$ is inspired by the computation of the top partner candidate's mass in the lattice~\cite{Ayyar:2018zuk} for a SU(4) gauge theory with 4 Weyl fermions in the anti-symmetric and another 4 in the fundamental representations of SU(4), which is intended to represent the candidate model 
 based on the SU(4) gauge theory with 5 EW charged Weyl fermions in the anti-symmetric and 3 QCD charged in each fundamental and anti-fundamental representations of SU(4)~\cite{Ferretti:2016upr}. 
Besides the top partner mass, this work provides the mass of the heavy gluon state $m_\rho \approx 6f$ which serves as cutoff of the dipole approximation, as well as the EW charged vector state with mass $6.5 f$ and the masses of other baryonic states.

In the plots of \fig{fig:Geff-pure} the dashed red line represents the point-like behavior while the solid blue curve is the behavior of $|G_{{\rm eff},T}|^2$ computed in the EFT limit where only operators up to dimension-6 operators have been considered, namely the first term in the expansion of the dipole expression. Notice that beyond $\Lambda_c$ the EFT expansion breaks down and the full form factor has to be considered. 

In each plot of \fig{fig:Geff-pure}, the leftmost arrow indicates the compositeness scale $\Lambda_{c}$, the middle arrow indicates the value of the lightest resonance $m_\rho=6f$ and the rightmost arrow indicates the kinematically accessible region for the production of heavy top pair $2m_T$.

It is interesting to note that the production cross section of a composite top partner is expected to be suppressed w.r.t. its point-like version, which  is usually considered in collider searches (see \emph{e.g.}~\cite{Buchkremer:2013bha,Aguilar-Saavedra:2013qpa,Benbrik:2019zdp}). In the pure composite case (left plot of \fig{fig:Geff-pure}) the suppression grows with energy, while for the partially composite state (right plot of \fig{fig:Geff-pure}), it stabilizes to a constant value because its elementary component dominates. The suppression is quite sizable for the specific values of the parameters we have considered but its magnitude can vary by changing the relative sizes of $M_D$ and $M$.
This suppression is expected for gluon initiated process too, as we are going to see in \sec{sec:gg}.
The mass $m_T$ we consider in this lattice motivated benchmark is beyond the reach of the LHC, even in the point-like top partner case with no suppression, but more optimistic scenario could be explored.

The unphysical peak at $s=4M^2$ in the right panel of \fig{fig:Geff-pure} is a consequence of the singular behavior of the Sachs form factors as discussed in \sec{sec:ff}. It is expected to be removed by the resonance contributions and do not alter the high energy behavior. Notice that for the pure composite state (left panel) the peak is not present because $G_{eff}$ in this case depends on a combination of $|G_E|^2$ and $|G_M|^2$ which is not singular.

%%%%%%%%%%%
\subsection{Probing the form factors in $q \bar q \to t\bar t$}
%%%%%%%%
Let us now consider the quark initiated top pair production $q \bar{q}\to g^*\to t\bar{t}$ which is a process that probes the top-quark form factors in the time-like region at energies $E$ such that $E \geq 2m_t$. 
The tree-level cross section for this process can be computed using the 
current in eq.~\eqref{tJ} and is given by
\be
\sigma_{q\bar{q}\to t\,\bar{t}}=\frac{8\alpha_{s}^{2}\pi}{27s}\sqrt{1-\frac{4m_{t}^{2}}{s}}\left(1+\frac{2m_{t}^{2}}{s}\right)|G_{\rm eff}(s)|^2
\label{eq:qqtoptop}
\ee
where 
\bea
|G_{\rm eff}(s)|^2&=&1+\left( 1+\frac{2m_t^2}{s}\right)^{-1}\Bigg\{%\left( 1+\frac{2m_t^2}{s}\right)+
\frac{2m_t^2+s}{M^2}(s_R^2+s_L^2){\rm Re}f_1(s) -\frac{3m_t}{M}s_Rs_L{\rm Re}F_2(s)\nn\\
&&+
\frac{s^2}{2M^4}\left[(s_R^4+s_L^4)\left(1-\frac{m_t^2}{s}\right)+6s_R^2s_L^2\frac{m_t^2}{s}\right]|f_1(s)|^2+\frac{8m_t^2+s}{8M^2}s_R^2s_L^2|F_2(s)|^2\nn\\
&&-\frac{3m_t s}{2M^3}s_Rs_L(s_R^2+s_L^2){\rm Re}f_1(s)F^*_2(s)\Bigg\}\,.
\label{eq:Geff}
\eea
Notice that in the limit of point-like heavy top partner we have that $f_1,F_2\to 0$ and $G_{\rm eff}(s)\to 1$. Similarly, if there is no mixing, namely $s_L,s_R\to 0$, the top-pair production cross section of \eq{eq:qqTT} would also behave as point-like.

\begin{figure}
\includegraphics[width=0.49\textwidth]{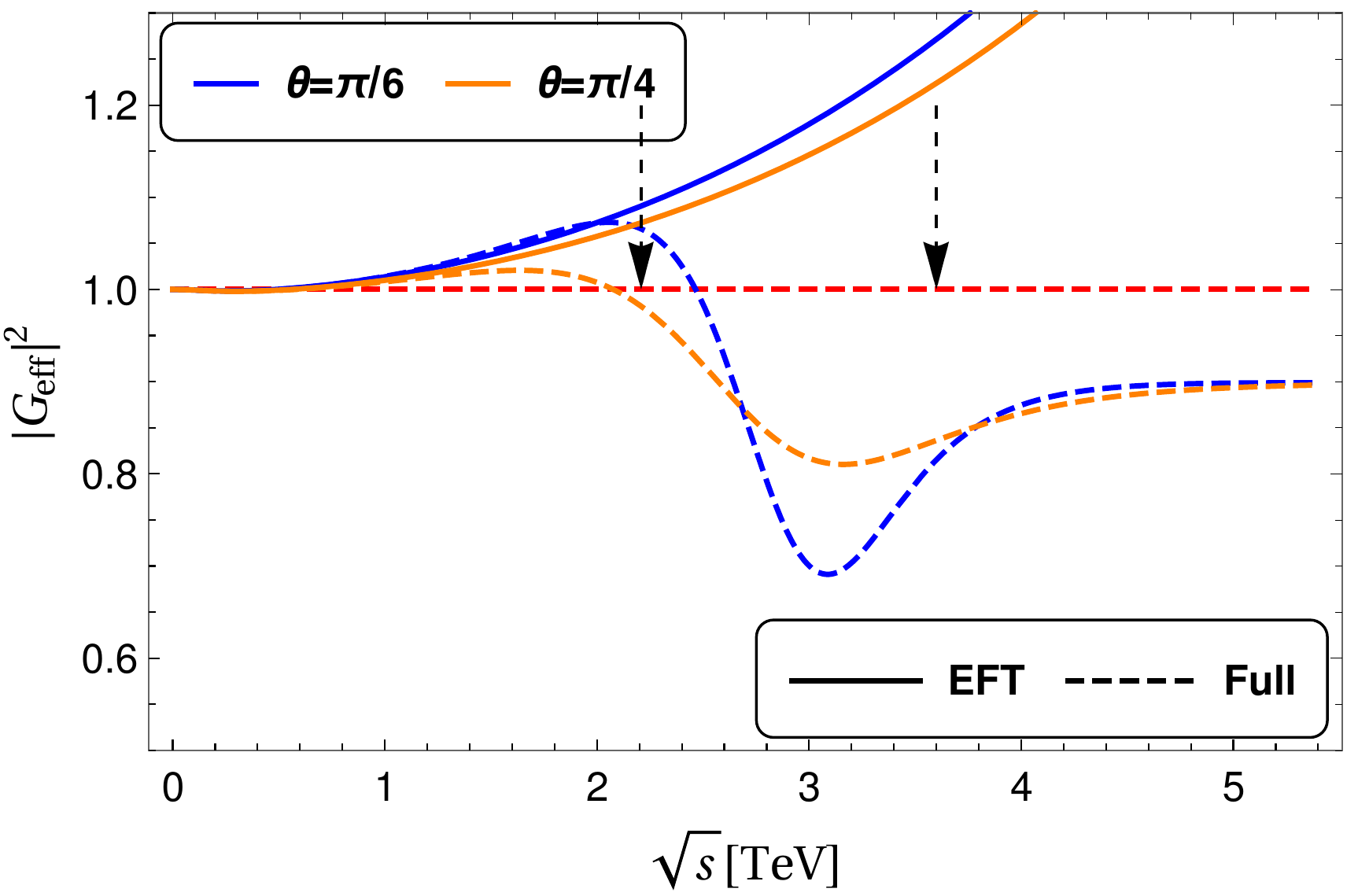}
\caption{$|G_{eff}|^2$. The  parameters are: $M=9f$, $M_D=5 f$, $f=0.6\TeV$, $\kappa_g=2$ and $\lambda=3$ ($s_L=0.091$ and $s_R=0.313$). The two arrows correspond from left to right to $\Lambda_{c}$ and $m_\rho=6f$. }
\label{fig:Geff}
\end{figure}

Fig.~\ref{fig:Geff} shows $|G_{eff}|^2$ as a function of the center of mass energy for two values of the absorptive phase $\eta=\pi/6$ and $\eta=\pi/4$ and the other parameters fixed as in \fig{fig:Geff-pure}, \emph{i.e.} $M_D=5f$, $M=9 f$, $\kappa_g=2$. 
The mixing angles are fixed by choosing $\lambda=3$, corresponding to $s_L=0.091$ and $s_R=0.313$. The mixing angles lead  to a large suppression of the Wilson coefficients.
The first arrow from the left in fig.~\ref{fig:Geff} indicates the values of $\Lambda_{c}$ while the second indicates the value of $m_\rho$ taken from the lattice (and represents the scale where the dipole approximation breaks down). 
The \emph{wiggle} around $M_D$ is a consequence of the mixing.%%%%%%%
%%%%%%%%%%%%%%%%%%

We can see that the full form factor can give peculiar signature not described by EFT or resonances. These effects can be searched as high energy deviations at collider experiments.

\begin{figure}
\includegraphics[width=0.49\textwidth]{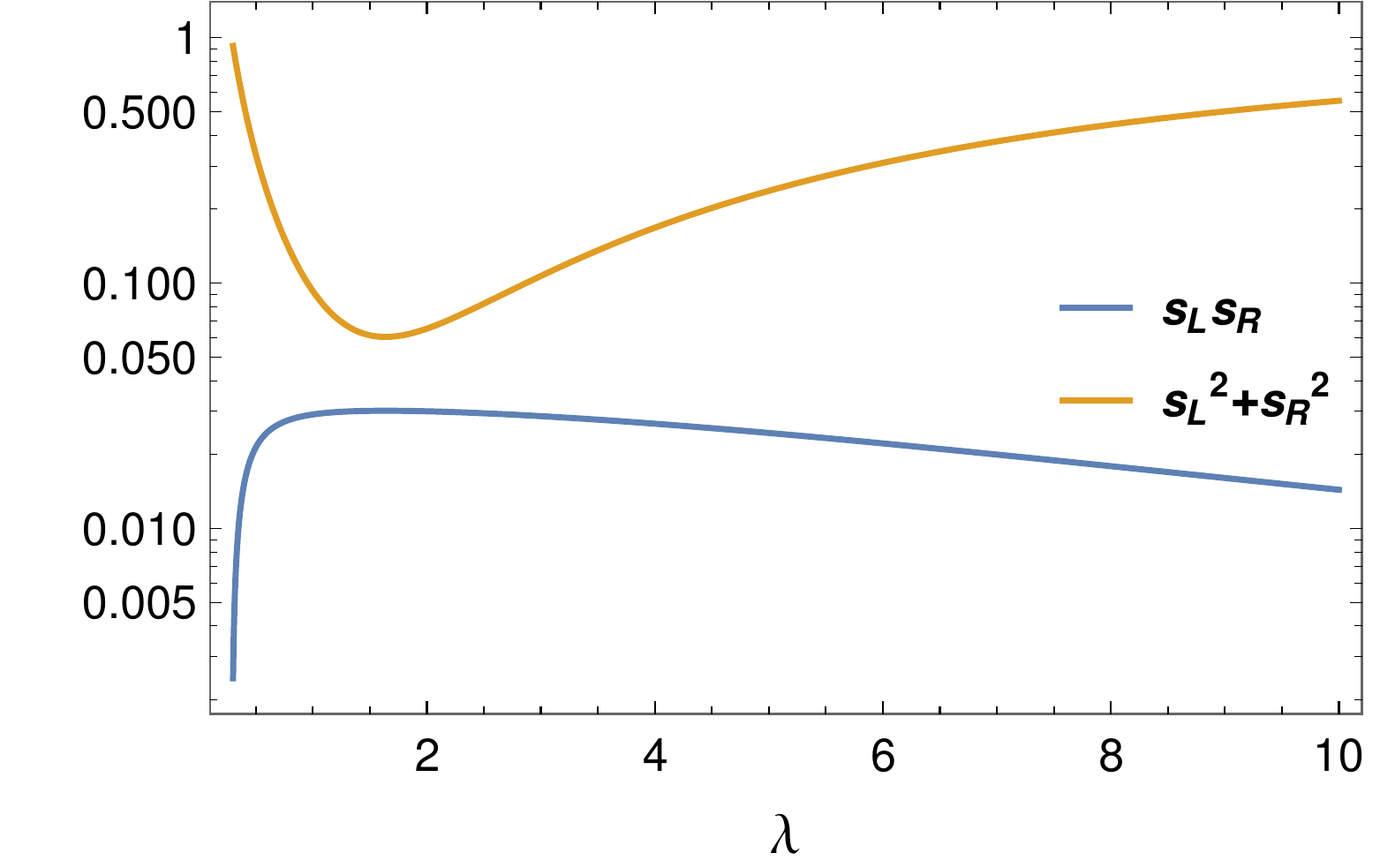}
\caption{Mixing angle dependence of the form factors $f_1$ ($s_L^2+s_R^2$) and $F_2$ ($s_L s_R$).}
\label{fig:sLsR}
\end{figure}

It is interesting to notice that the main compositeness effect in $qq\to g^*\to t\bar{t}$ at low and intermediate scales comes from the Dirac-like interaction. 
This fact originates from two observations. First, as shown in \eq{ltopeff}, the chromomagnetic moment is suppressed by the mixing angle combination $s_L s_R$, which is typically much lower than the combination $\sim s_L^2+s_R^2$ of the Dirac-like interaction. Therefore, although the typical compositeness scales are very similar for both operators, the chromomagnetic dipole moment operator is more suppressed by the overall mixing angle factor. To illustrate this fact we show in \fig{fig:sLsR} the values of $s_L s_R$ and $s_L^2+s_R^2$ as a function of $\lambda$ for the specific scenario described in \sec{sec:PC}. 
Second, the chromomagnetic interaction has different helicity compared to the point-like interaction; therefore, the interference between the two terms (which dominates for small values of the leading coefficient) carries a suppression of order $\mathcal{O}(m_t/\sqrt{s})$, which cancels the naive expectation of energy growing behavior of the linear EFT term. This fact can be explicitly seen in \eq{eq:Geff}.

For this choice of parameters, with respect to SM expectations, the presence of form factors would yield an enhancement of the total cross section in $q\bar{q}$ initiated top quark pair production of $\sim 1\%$ at the LHC with $\sqrt{s}=13\TeV$. We computed the total cross section by convoluting of the partonic cross section with the CTEQ5 parton distribution function~\cite{Lai:1999wy}.
Imposing a cut on the invariant mass of the top pair $m(t\bar t)>1\TeV$ we get an enhancement of 4\%.
Such small effects at the differential level could in principle be detected with a careful study, as illustrated by the recent measurement of charge asymmetry in $t\bar{t}$ production~\cite{ATLAS:2019czt}.

%%%%%%%%%%%
\section{Two gluons phenomenological form factors}
\label{sec:gg}
%%%%%%%%%%%

In this section we consider the gluon-initiated processes $gg\to T\bar T$ and $gg\to t\bar t$. 
Assuming $T$ to be a pure composite state, we can borrow the knowledge about the nucleon EM interactions and follow a similar phenomenological prescription to treat $gg$-induced $T$ pair production. The analog to $gg\to T\bar T$ in the nucleon case is the production of proton-antiproton via photon scattering $\gamma\gamma\to p\bar p$, which has been modelled and observed in collider experiments~\cite{Poppe:1986dq,Klusek-Gawenda:2017lgt}. 
In this section we extend the framework of ~\cite{Klusek-Gawenda:2017lgt} to treat these gluon initiated processes, \emph{i.e.} top and top-partner pair production.

\subsection{Top partner production via $gg\to T\bar T$}

Our phenomenological approach is based on Ref.~\cite{Klusek-Gawenda:2017lgt}, where the scattering amplitudes of the process $\gamma\gamma\to p\bar p$ are estimated from the sum of three different contributions: 1- the proton exchange in the $t$ and $u$-channels diagrams, 2- the exchange of mesons in the $s$-channel and 3- the so-called hand-bag mechanism. The resulting amplitudes give a good fit to data of several experiments. 
 The only resonance that give relevant contribution to the amplitude is $f_2(1950)$ which is very close to the production threshold energies $E=2m_p\sim 1876$ MeV. The lighter state $f_2(1270)$ gives only marginal contribution orders of magnitudes smaller. At larger energies $E\gtrsim 3.3\GeV$ the handbag mechanism that probes the proton contituents take over and dominates the scattering amplitude.

In our extrapolation of this method to the composite top sector, we will neglect the resonance and the handbag contributions. 
We assume for simplicity the absence of resonances near the threshold production. The handbag on the other hand becomes relevant only at much higher energies and can be neglected in our exploratory study. 

The basic ingredients of the proton-exchange calculation in \cite{Klusek-Gawenda:2017lgt} are the proton exchange amplitudes computed using the general proton-photon interaction vertex of  \eq{emff} and an overall form factor that multiplies the whole amplitude that parametrizes the effect of proton off-shell-ness, while keeping gauge invariance and crossing symmetry. 

Let us first consider the case of a purely-composite $T$. In addition to the analog of the proton-exchange diagrams, computed by using the one-gluon interaction vertex of \eq{TprimeJ}, one needs to consider extra diagrams contributing to $g g\to T\bar{T}$ that are present because of the non-abelian nature of the strong interactions. In order to preserve gauge invariance, we use the effective Lagrangian in \eq{TprimeEA} to derive the Feynman rules and compute the $g g\to T\bar{T}$ scattering amplitude which involve three family of diagrams: $t$- and $u$-channel $T$-exchange diagrams, $s$-channel gluon-exchange diagrams and $g g T\bar{T}$ contact interaction diagrams. We used FeynArts and FormCalc~\cite{Hahn:2000kx,Hahn:2000jm} to compute the tree level $g g\to T\bar{T}$ amplitude $\mathcal{M}_{T,bare}$ which we next multiply by an overall form factor to get the final result
\begin{equation}
\mathcal{M}_T = F(t,u,s)\mathcal{M}_{T,bare}\,,
\label{eq:ggampff}
\end{equation}
with
\begin{equation}\label{eq:ggff}
F(t,u,s)=\frac{\hat F(t,u,s)^2+\hat F(u,t,s)^2}{1+\tilde{F}(t,u,s)^2}
\end{equation}
and 
\begin{eqnarray}
\hat F(t,u,s)&=&\exp\left( -\frac{s+u-t}{2\Lambda_T^2}\right) \,,\quad
\tilde F(t,u,s)=\exp\left( \frac{s+2t+2u}{\Lambda_T^2}\right) \,.
\end{eqnarray}
The specific form of $F(t,u,s)$ is taken from \cite{Klusek-Gawenda:2017lgt}.

The case of the partially composite heavy $T$ quark is a bit more involved. We therefore consider first the bare amplitude $\mathcal{M}_{T,bare}$ for $gg\to T\bar T$ in the presence of form factors computed by using the lagrangian in eq.~\eqref{LefftT}. We have that
\be 
\mathcal{M}_{T,bare}=\mathcal{M}_{T,0}+\mathcal{M}_{T,F_2}\,,
\ee
where $\mathcal{M}_{T,0}$ is the point-like amplitude for $gg\to T\bar T$ and $\mathcal{M}_{T,F_2}$ is the $F_2$ dependent part ($f_1$ does not contribute to the amplitude). In order to take into account the partially composite nature of the top-quark, while preserving gauge invariance, we parametrize the final amplitude for the process $gg\to T\bar T$ as follows
\be 
\mathcal{M}_T =\frac{s_L^2+s_R^2}{2}\mathcal{M}_{T,0}+F(t,u,s)\left( \frac{c_L^2+c_R^2}{2}\mathcal{M}_{T,0}+\mathcal{M}_{T,F_2}
\right)\,,
\label{eq:gg2TT}
\ee
where $F(t,u,s)$ is given in eq.~\eqref{eq:ggff}. 

This \emph{ansatz} for the amplitude has the property of recovering the two extreme cases: for $c_{L,R}\to 0$ we recover the point like case (because also $\mathcal{M}_{T,F_2}$ vanishes in this limit), whereas for $c_{L,R}\to 1$ we recover the purely composite case of eq.~\eqref{eq:ggampff}. Notice that if we had used in the partially composite case $F(t,u,s)$ as overall form factor for $\mathcal{M}_{T,bare} $ as in eq. \eqref{eq:ggampff}, then the cross section would have been suppressed also in the point-like limit. 

\begin{figure}
\includegraphics[width=0.49\textwidth]{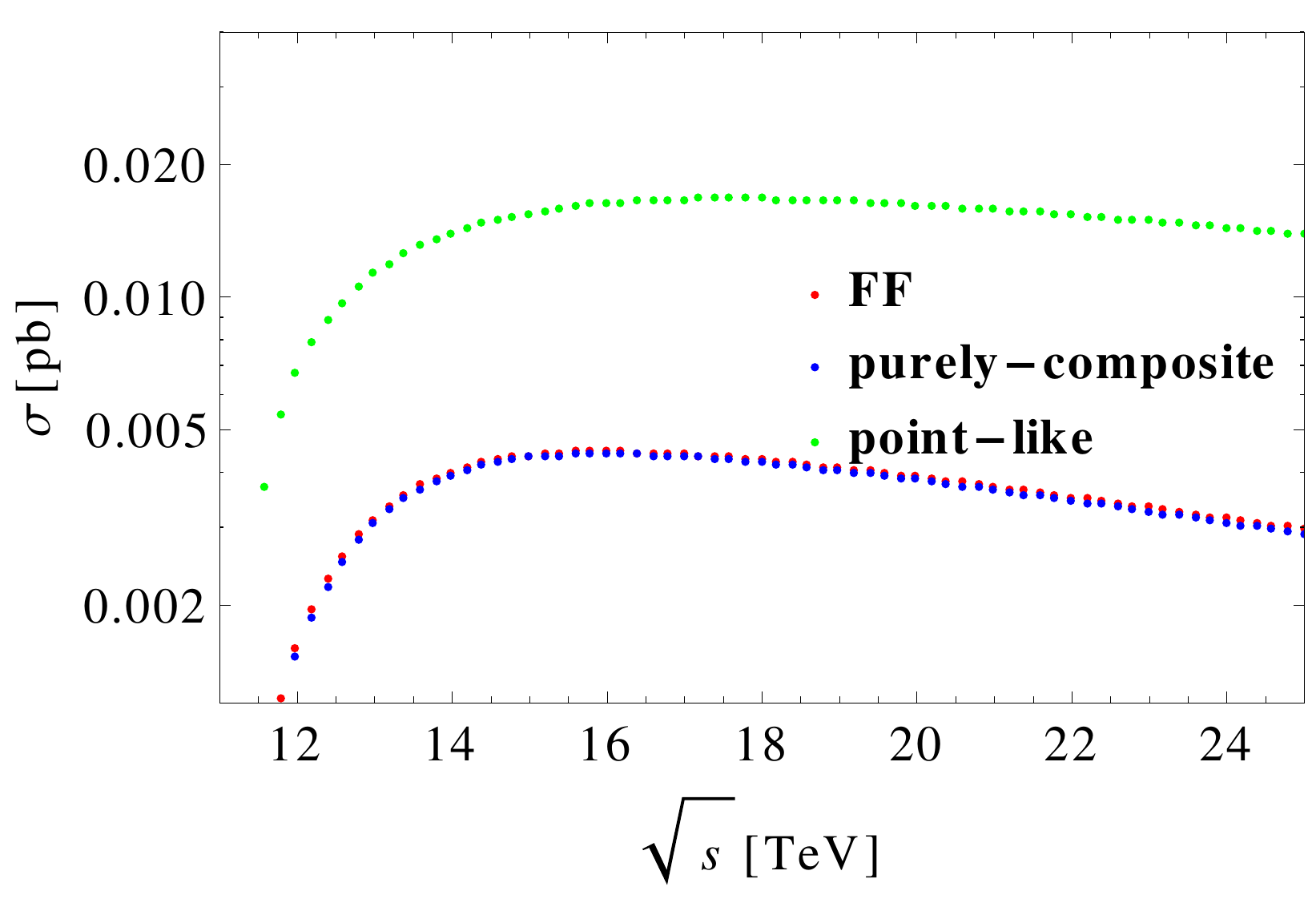}
\caption{Cross section for $gg\to T\bar{T}$ as function of center of mass energy in the partially composite case (red). Purely-composite (blue) and point-like (green) cases are also shown for comparison. Values of parameter used: $\lambda=3$, $f=0.6\TeV$, $\kappa=2$, $m_D=5f$, $M=9f$ and $\Lambda_T=11f$. }
\label{fig:gg-TT}
\end{figure}

Using the parameters of the previous section ($\lambda=3$, $f=0.6\TeV$, $\kappa=2$, $m_D=5f$, $M=9f$) and fixing $\Lambda_T=11f$, which is inspired by the proton case for where $\Lambda_p\gtrsim m_p$ (see \app{app:proton} for more detail), we compute the total cross section for $gg\to T\bar T$ as function of the center of mass energy and the result is shown in Fig.~\ref{fig:gg-TT}. In the same plot we compare with the purely-composite and point-like cases.

We can notice a large suppression of the cross section with respect to the point-like case. Current searches are performed under the assumption of a point-like top partner. 
We consider a point-like scenario hard to be realized in a composite framework, even for a light top partner, in particular due to the anomalous chromo-magnetic moment that should be present even for high compositeness scales. 
Another expectation is the raise of inelastic processes over the  highly suppressed  elastic cross section. Only in scenario where other processes are kinematically inaccessible, the top-partner production can be relevant.

%%%% 
\subsection{Top-quark pair production via $gg\to t\bar t$}
%%%%%%%
To describe top pair production we follow exactly the same principles.
We compute the bare amplitude $\mathcal{M}_{t,bare}$ for $gg\to t\bar t$ in the presence of form factors by using the lagrangian in eq.~\eqref{LefftT}. We have that
\be 
\mathcal{M}_{t,bare}=\mathcal{M}_{t,0}+\mathcal{M}_{t,F_2}
\ee
where $\mathcal{M}_{t,0}$ is the SM amplitude for $gg\to t\bar t$ and $\mathcal{M}_{t,F_2}$ is the $F_2$ dependent part. 
Then, analougously to \eq{eq:gg2TT} we parametrize the amplitude for the process $gg\to t\bar t$ as follows
\be 
\mathcal{M}_t =\frac{c_L^2+c_R^2}{2}\mathcal{M}_{t,0}+F(t,u,s)\left( \frac{s_L^2+s_R^2}{2}\mathcal{M}_{t,0}+\mathcal{M}_{t,F_2}
\right)\,.
\ee
As already mentioned, this \emph{ansatz} reproduces the correct limits.
Despite its arbitrariness, we will see that the particular form of this parametrization does not change our conclusions.

\begin{figure}
\includegraphics[width=0.49\textwidth]{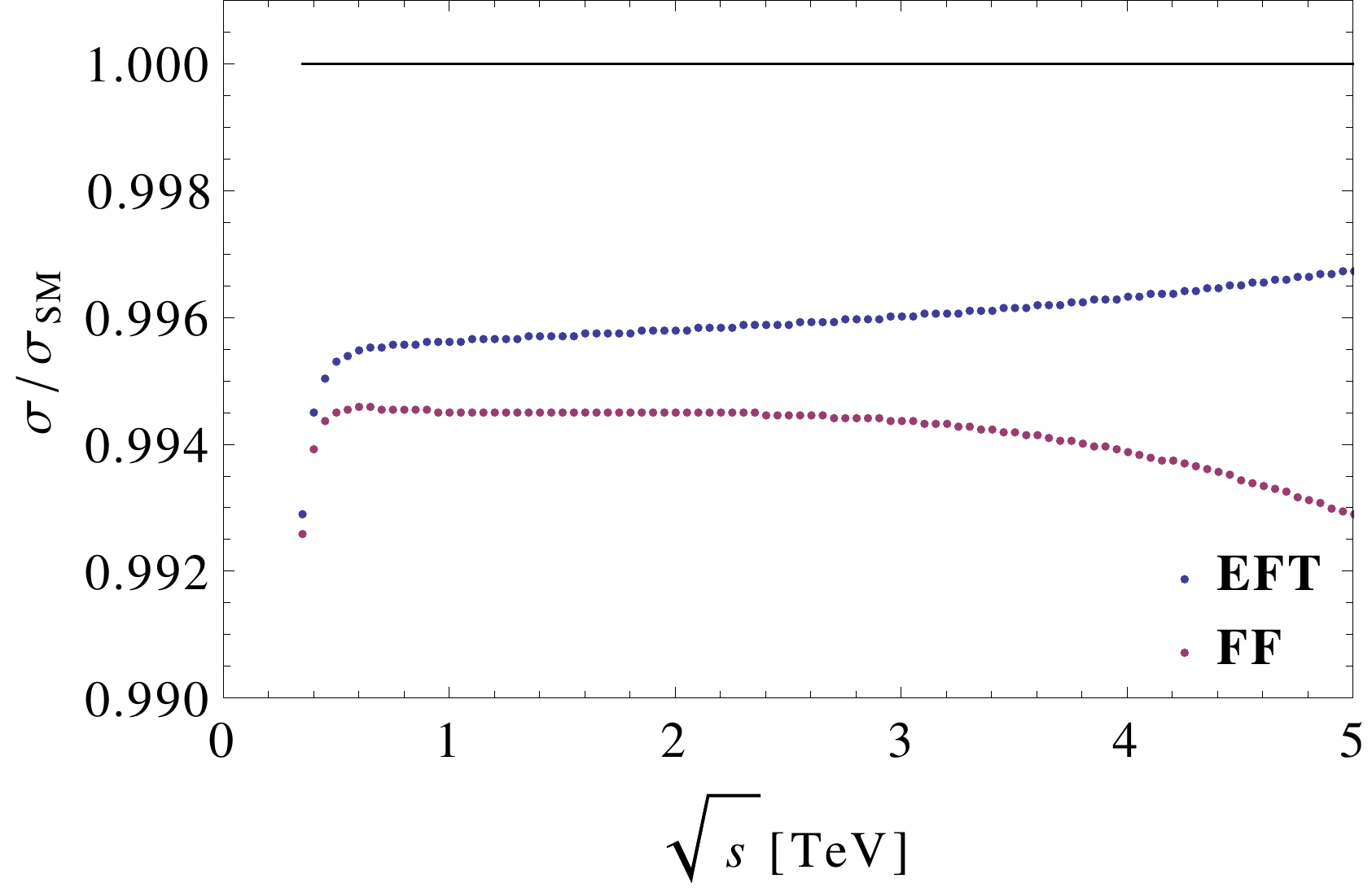}
\caption{Cross section for $gg\to t\bar{t}$ normalized to the SM as function of center of mass energy in the partially composite case (purple). The EFT case (blue) is also shown for comparison. Values of parameter used: $\lambda=3$, $f=0.6\TeV$, $\kappa=2$, $m_D=5f$, $M=9f$ and $\Lambda_T=11f$.}
\label{fig:gg-toptop}
\end{figure}

The resulting total cross section as function of center of mass energy normalized to the SM (point-like) prediction is shown in \fig{fig:gg-toptop}. 
For comparison, we show the same ratio for the leading EFT term.

It can be noticed that the form factor relative effect can indeed be larger than EFT as the energy grow. However, the overall effect is expected to be small.
This fact can be understood by two observations.
First, by noticing that the Dirac-type interactions described by $f_1$ vanish identically at tree-level in this process ($gg\to t\bar t$). 
The vanishing of this contribution can be understood in the language of EFT: the operator in the second term of \eq{ltopeff} that leads to the Dirac-like interaction is equivalent by the EOM to several four-fermion interactions that do not contribute to $gg\to t\bar t$ at tree level.
Second, similarly to the quark-initiated case, the Pauli-type interaction, which is the only contribution to the process, is doubly suppressed, by the small product of mixing angles $s_L s_R$ and by the helicity flip w.r.t. the SM leading behavior in energy.

%%%%%%%%%%%%
\section{Conclusion}
\label{sec:conclusion}
%%%%%%%%%%%%

In this work we have parametrized the gluonic interactions of an hypothetical partially composite top-quark via two form factors, beyond the leading dimension EFT prescription. 
We argued that, in the case where the compositeness scale is accessible, the full energy dependence of these interactions cannot be neglected. 
We discussed the effects of form factors in $q\bar{q}$-initiated top and heavy-top pair production. For that purpose, we considered a simple phenomenological parametrization of one-gluon form factors, based on the proton 
dipole approximation. Furthermore, we expanded this prescription to $gg$-initiated processes, using as guidance the modeling of proton-antiproton production in photon-photon scattering.

We showed that the $q\bar{q}$-initiated top pair production is more sensitive to the structure of the top-quark than the $gg$-initiated process. This observation comes from the fact that the operators associated with the form factor $f_1(q^2)$ does not contribute to the $gg$-initiated process and the chromomagnetic moment is doubly suppressed. 
In our phenomenological model, the effect in $q\bar{q}\to t\bar t$ would appear as a \emph{wiggle} in the invariant mass distribution of the top-quark pair system that could be in principle looked for at high energy and high luminosity collider experiments. 
Further studies must be carried out to assess the LHC potential in that matter. 
It is interesting to notice that small charge asymmetry effects from quark initiated top pair production  have been recently measured with good accuracy by the ATLAS collaboration~\cite{ATLAS:2019czt}.
Moreover, we showed that the production of a composite top partner is expected to be  suppressed compared to a point-like state, which is typically considered as benchmark model for LHC top partner searches. This fact should be taken into account in future searches.
In addition, we pointed out the existence of a new form of single top-partner production via QCD interaction $pp\to t\bar{T},T\bar{t}$ that must be further investigated.

\section*{Acknowledgements}

We would like to thank Jose Llanes for his contributions in the early stages of this work. 
 DBF acknowledges financial support from the Knut och Alice Wallenberg foundation under the grant KAW 2017.0100 (SHIFT project).
 DBF would like to thank Gabriele Ferretti for reading the manuscript and giving valuable comments.

\appendix

\section{Derivation of the most general interaction current and form factors}
\label{app:genericFF}
%%%%%%%%%%%%%%%%%%%%%%
Consider a fermion state $\psi$ with mass $m$, then the most general Lorentz invariant and  parity conserving one-gluon interaction current can be written as
\be 
J^\mu_a=g_{s}\bar{\psi}(p) T_{a}\Gamma^\mu(p,p') \psi(p')\,
\ee
where the interaction vertex is
\begin{equation}
\Gamma^\mu(p,p')=\Gamma^\mu(q,m)=A(q^2,m)\gamma^\mu-\frac{t^\mu}{\Lambda}B(q^2,m)+\frac{q^\mu}{\Lambda}C(q^2,m)\,,
\end{equation}
and
\be 
q=p-p'\qquad {\rm and} \qquad t=p+p'\,.
\ee
The form factors $A$,$B$ and $C$ can be taken without loss of generality to be scalar functions\footnote{They could involve Dirac matrices dotted into vectors, namely $\slashed p$ or $\slashed p'$. However, these terms can be rearranged and transformed into ordinary numbers, depending on $m$, via the Dirac equation. } of $q^2$ and $m$. Moreover, they can depend on some  intrinsic compositeness scale related to the strong dynamics that we denote by $\Lambda$. We will not write explicitly this dependence on $\Lambda$, in order to keep the notation as light as possible.

The QCD Ward identity 
\be 
 \bar{\psi}(p) T_{a}\,q_\mu\Gamma^\mu(q,m)\, \psi(p')=0
\ee
requires that 
\begin{equation}
C(q^2,m)=0\,.
\end{equation}
Therefore the most general gauge invariant interaction vertex reduces to
\begin{equation}
\Gamma^\mu(q,m)=A(q^2,m)\gamma^\mu-\frac{t^\mu}{\Lambda}B(q^2,m)\,.
\end{equation}
We can finally use the Gordon identity to rewrite this vertex in the standard form
\begin{equation}
\Gamma^\mu(q,m) = \left[A(q^2,m) - \frac{2 m }{\Lambda}B(q^2,m)\right]\gamma^\mu+i \frac{\sigma^{\mu\nu}q_\nu}{\Lambda}B(q^2,m)\,,
\end{equation}
Defining 
\begin{eqnarray}
F_1(q^2,m) &=& A(q^2,m) - \frac{2 m }{\Lambda}B(q^2,m) \\
F_2(q^2,m) &=& 2B(q^2,m)\,,
\end{eqnarray}
we get the more familiar expression for the vertex
\begin{equation}
\Gamma^\mu(q,m)=F_1(q^2,m)\gamma^\mu + \frac{i\sigma^{\mu\nu}q_{\nu}}{2\Lambda}F_{2}(q^2,m)\,.
\end{equation}
In order to have the correct QCD charge normalization we need that $F_1(0,m) =1$, therefore we can write
\be 
F_1(q^2,m)=1+\frac{q^2}{\Lambda^2}f_1(q^2,m)\,.
\ee
It is reasonable to assume that these form factors will depend negligibly on the mass of the fermionic state. The idea is that the main dependence comes just from the probed gluon momenta $q^2$ (and from the intrinsic scale $\Lambda$ related to the strong dynamics), namely
\be 
F_i=F_i(q^2).\label{eq:assumptionFF}
\ee
The following heuristic arguments to justify this assumption can be provided : 1 - the form factor should not depend on the kinematic mass of the incoming fermion but only on the probing momenta $q$; 2 - if the masses entering in the expressions of the form factors are of kinematical origin, i.e. $M=m$ in  \eq{F12dapprox} then the typical compositeness scale in \eq{eq:lambdac} is dominated by the spinor mass $m$, which for the top quark would be too small. And 3 - the effective lagrangian we use to derive the interaction (see Appendix B) is motivated by modifications that originate in the gauge sector and do not ``affect" the fermion fields. That Lagrangian gives only $q^2$ dependent form factors.

We can now consider first the pure composite state $T'$ in absence of mixing and write the following interaction current
\be \label{TprimeJapp}
(J_{T'})^\mu_a=g_{s}\bar{T'} T_{a}\left[\gamma^{\mu}F_{1}(q^{2})+\frac{i\sigma^{\mu\nu}q_{\nu}}{2M}F_{2}(q^{2})\right]T'\,,
\ee
where $F_1(q^2)=1+\frac{q^2}{M^2}f_1(q^2)$ and $F_2(q^2)$ are the Dirac and Pauli form factors, respectively. The mass parameter $M$ entering in \eq{TprimeJapp} is understood to be related to the strong dynamics scale, namely $M=M(\Lambda)$.

Now let us switch on the mixing terms. Under the assumption of \eq{eq:assumptionFF}, the form factors for the mass eigenstates $t$ and $T$ can be derived straightforwardly by performing the appropriate rotation to the mass basis 
\begin{eqnarray}
F_1^t(q^{2})&=&1+(s_L^2P_L+s_R^2P_R)\frac{q^2}{M^2}f_1(q^{2})\\
F_2^t(q^{2})&=& -s_L s_R F_2(q^{2}) \\
F_1^T(q^{2})&=&1+(c_L^2P_L+c_R^2P_R)\frac{q^2}{M^2}f_1(q^{2})\\
F_2^T(q^{2})&=&c_L c_R F_2(q^{2})\,.
\label{eq:F1toff}
\end{eqnarray}

%%%%%%%%%%%%%%%%%
\section{Effective lagrangian of form factors}
\label{app:gaugeinvff}
%%%%%%%%%%%%%%%%%%%%%%
The form of the current in \eq{TprimeJapp} can be derived from the following gauge invariant effective lagrangian
\begin{eqnarray}\label{TprimeEA}
{\cal L}_{T'} & = & \bar{T}'i\gamma^{\mu}D_{\mu}T'+\frac{g_{s}}{M^2}\bar{T}'\gamma^{\mu}T_{a}T'f_{1}(-D^{2})D^{\nu}G_{\mu\nu}^{a}+\frac{g_{s}}{4M}\bar{T}'\sigma^{\mu\nu}T_{a}T'F_{2}(-D^{2})G_{\mu\nu}^{a}\nn\\ 
\end{eqnarray}
where $D_{\mu}T'=(\partial_{\mu}-ig_{s}G_{\mu})T'$ and $D^{\nu}G_{\mu\nu}^{a}=\partial^{\nu}G_{\mu\nu}^{a}+g_{s}f^{a}{}_{bc}G^{\nu b}G_{\mu\nu}^{c}$. The form factors $f_1$ and $F_2$ are functions of the covariant laplacial $D^2$ in order to ensure gauge invariance. Deriving the Feynman rule for the $\bar T' T' g$ vertex is it possible to show that the relation between the form factor $f_1$ appearing in eq.~\eqref{TprimeEA} and $F_1$ of eq.~\eqref{TprimeJapp} is given by 
\be 
F_1(q^2)=1+f_1(q^2)\frac{q^2}{M^2}\,.
\ee
The elementary top $t'$ on the other hand has the following lagrangian
\be 
{\cal L}_{t'} = \bar{t}'i\gamma^{\mu}D_{\mu}t'
\ee
therefore the total effective lagrangian involving $t'$ and $T'$ is given by
\be 
{\cal L}={\cal L}_{T'}+{\cal L}_{t'}+{\cal L}_{\rm mass}
\ee
where ${\cal L}_{\rm mass}$ is given in eq.~\eqref{Lmass}. After diagonalization by means of eq.~\eqref{Ttdiagon} we obtain the following lagrangian
\begin{eqnarray}
 \label{LefftT}
{\cal L}&=&\bar{T}i\gamma^{\mu}D_{\mu}T+\bar{t}i\gamma^{\mu}D_{\mu}t-m_T \bar T T -m_t \bar t t\nonumber\\
&&+\frac{g_{s}}{M^2}\bar{t}\gamma^{\mu}T_{a}(s_L^2 P_L + s_R^2 P_R)tf_{1}(-D^{2})D^{\nu}G_{\mu\nu}^{a}\nn\\
&& - \frac{g_{s}}{4M}s_Ls_R\bar{t}\sigma^{\mu\nu}T_{a}tF_{2}(-D^{2})G_{\mu\nu}^{a}\nonumber\\
&&+\frac{g_{s}}{M^2}\bar{T}\gamma^{\mu}T_{a}(c_L^2 P_L + c_R^2 P_R)T\, f_{1}(-D^{2})D^{\nu}G_{\mu\nu}^{a}\nn\\
&& + \frac{g_{s}}{4M}c_Lc_R\bar{T}\sigma^{\mu\nu}T_{a}T\,F_{2}(-D^{2})G_{\mu\nu}^{a}\nonumber\\
&&+\frac{g_{s}}{M^2}\bar{t}\gamma^{\mu}T_{a}(s_Rc_R P_R - s_L c_L P_L)T\, f_{1}(-D^{2})D^{\nu}G_{\mu\nu}^{a} + (t\leftrightarrow T) \nonumber \\
&&+ \frac{g_{s}}{4M}\bar{t}\sigma^{\mu\nu}T_{a}(-s_Lc_RP_R+s_Rc_LP_L)T\,F_{2}(-D^{2})G_{\mu\nu}^{a}\nn\\
&& + \frac{g_{s}}{4M}\bar{T}\sigma^{\mu\nu}T_{a}(c_Ls_RP_R-c_Rs_LP_L)t\,F_{2}(-D^{2})G_{\mu\nu}^{a}
\end{eqnarray}
Thanks to the partial compositeness mechanism, the mixing of the elementary top with the composite top induces a modification in the interaction of the light top with the gluon. From the lagrangian in eq.~\eqref{LefftT} we can derive the interaction current of eq.~\eqref{tJ}.

Let us present here the Feynman rules involving only the top-quark $t$ which can be obtained from the lagrangian in eq.~\eqref{LefftT}. The Feynman rules for the three point function are
\be \label{eq:fr1}
t_1\, \bar{t}_2\,g_3\, :\,ig_s \gamma^{\mu_3}T_{a_3}+ V_1^g(a_3,\mu_3,p_3)f_1(p_3^2)+V_2^g(a_3,\mu_3,p_3)F_2(p_3^2)
\ee
where
\be 
V_1^g(a_3,\mu_3,p_3)=i\frac{g_s}{M^2} T_{a_3}(\gamma^{\mu_3}p_3^2-p_3^\mu\slashed{p_3})(s_L^2P_L+s_R^2P_R)
\ee
\be 
V_2^g(a_3,\mu_3,p_3)=-\frac{g_s}{2M}s_Ls_RT_{a_3}\sigma^{\mu_3\nu}p_{3\nu}\,.
\ee
The Feynman rules for the four point function are
\bea \label{eq:fr6}
t_1\, \bar{t}_2\,g_3\,g_4\, &:& V_1^{gg}f_1(p_{34}^2)+ig_sf^a{}_{a_3a_4}f'_1(p_{34}^2)\Bigg[
V_1^g(a,\mu_4,p_4)(p_3+2p_4)_{\mu_3}\nonumber\\
&&-V_1^g(a,\mu_3,p_3)(p_4+2p_3)_{\mu_4}\Bigg]+V_2^{gg}F_2(p_{34}^2)\nn\\
&&+ig_sf^a{}_{a_3a_4}F'_2(p_{34}^2)\Bigg[
V_2^g(a,\mu_4,p_4)(p_3+2p_4)_{\mu_3}
-V_2^g(a,\mu_3,p_3)(p_4+2p_3)_{\mu_4}\Bigg]+\ldots\nn\\
\eea
where
\be 
V_1^{gg}=\frac{g_s^2}{M^2}f^a{}_{a_3a_4}T_{a}\left[ \gamma^{\mu_3}(p_4+2p_3)_{\mu_4}-
\gamma^{\mu_4}(p_3+2p_4)_{\mu_3}-\eta^{\mu_3\mu_4}(\slashed{p_3}-\slashed{p_4})
(s_L^2P_L+s_R^2P_R)\right]
\ee
\be 
V_2^{gg}=-\frac{g_s^2}{4M}s_Ls_Rf^a{}_{a_3a_4}T_{a}( \gamma^{\mu_3} \gamma^{\mu_4}- \gamma^{\mu_4} \gamma^{\mu_3})
\ee
and the ellipses denote terms that vanish for on-shell gluons. In the expression above we have defined $p_{34}=p_3+p_4$ and
\begin{equation}
F'(s)=\frac{F(s)-F(0)}{s}\,.
\end{equation}
The Feynman rules involving the heavy-top $T$ can be derived from the lagrangian in eq.~\eqref{LefftT} in complete analogy.

\section{Proton form factors}
\label{app:proton}

In the model proposed in 1972~\cite{Iachello:1972nu} and  subsequentially improved~\cite{Iachello:2004aq,Bijker:2004yu}, the external photon interact both with an intrinsic structure 
\begin{equation}
g(q^2)=(1-\gamma e^{i\theta  \Theta(q^2)}q^2)^{-2}
\end{equation}
with $\Theta(s)$ the Heaviside function, which encodes the interaction with the constituents quarks and must reproduce the asymptotic behavior of perturbative QCD, and the interaction with a meson cloud, which can be approximated by exchange of vector meson resonances in the spirit of the Vector Meson Dominance. 
In this version of the model the form factors are given by 
\begin{eqnarray}
F_1^S(q^2)&=&\frac{1}{2}g(q^2)\left[(1-\beta_\omega-\beta_\varphi)+ \beta_\omega\frac{m_\omega^2}{m_\omega^2-q^2} + \beta_\varphi\frac{m_\varphi^2}{m_\varphi^2-q^2}\right] \\
F_1^V(q^2)&=&\frac{1}{2}g(q^2)\left[(1-\beta_\rho)+ \beta_\rho\frac{m_\rho^2}{m_\rho^2-q^2}\right] \\
F_2^S(q^2)&=&\frac{1}{2}g(q^2)\left[(\kappa_p+\kappa_n-\alpha_\varphi)\frac{m_\omega^2}{m_\omega^2-q^2} + \alpha_\varphi\frac{m_\varphi^2}{m_\varphi^2-q^2}\right] \\
F_2^V(q^2)&=&\frac{1}{2}g(q^2)\left[(\kappa_p-\kappa_n)\frac{m_\rho^2}{m_\rho^2-q^2}\right] 
\end{eqnarray} 
Notice that the $F_1$ functions behave asymptotically as $1/q^4$ while $F_2$ go with $1/q^6$.
The width effects can be incorporated by the modification of the propagator by
\begin{equation}
\frac{m_\rho}{m_\rho^2-q^2}\to \frac{m_\rho^2+8\Gamma_\rho m_\pi/\pi}{m_\rho^2-q^2+(4m_\pi^2-q^2)\Gamma_\rho\alpha(q^2)/m_\pi}
\label{eq:prop}
\end{equation}
with
\begin{equation}
\alpha(s)=\begin{cases}
\frac{2}{\pi}\sqrt{1-\frac{4m_\pi^2}{s}}\,\log\left(\frac{\sqrt{4m_\pi^2-s}+\sqrt{-s}}{2m_\pi}\right) \text{ if } s<0\\
\frac{2}{\pi}\sqrt{\frac{4m_\pi^2}{s}-1}\,\text{Arctan}\left[\left(\frac{4m_\pi^2}{s}-1\right)^{-1/2}\right] \text{ if } 0<s<4m_\pi^2\\
\frac{2}{\pi}\sqrt{1-\frac{4m_\pi^2}{s}}\,\log\left(\frac{\sqrt{s-4m_\pi^2}+\sqrt{s}}{2m_\pi}\right)-i\sqrt{1-\frac{4m_\pi^2}{s}} \text{ if } s>4m_\pi^2
\end{cases}
\end{equation}

\begin{figure}
\includegraphics[width=0.49\textwidth]{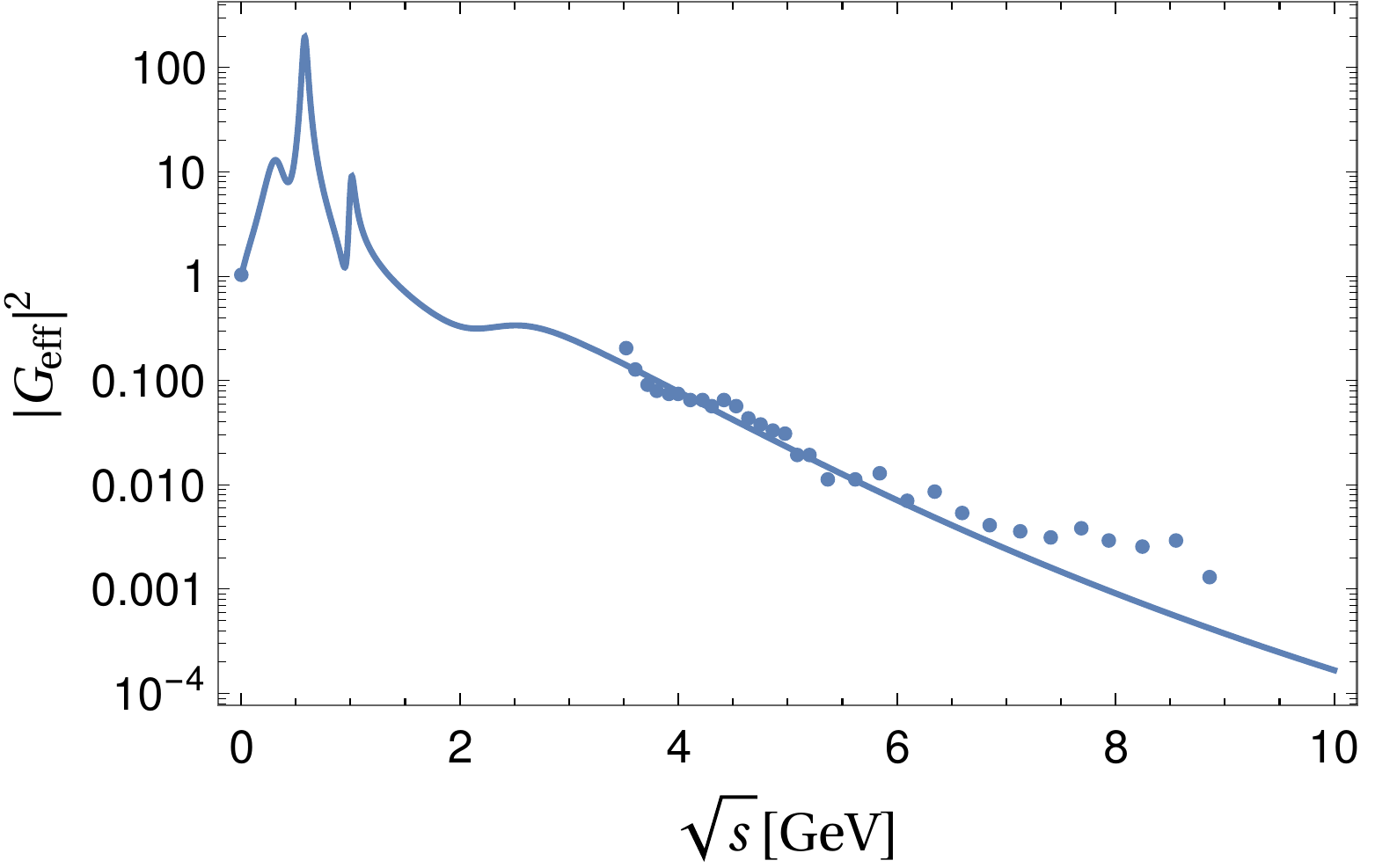}
\includegraphics[width=0.49\textwidth]{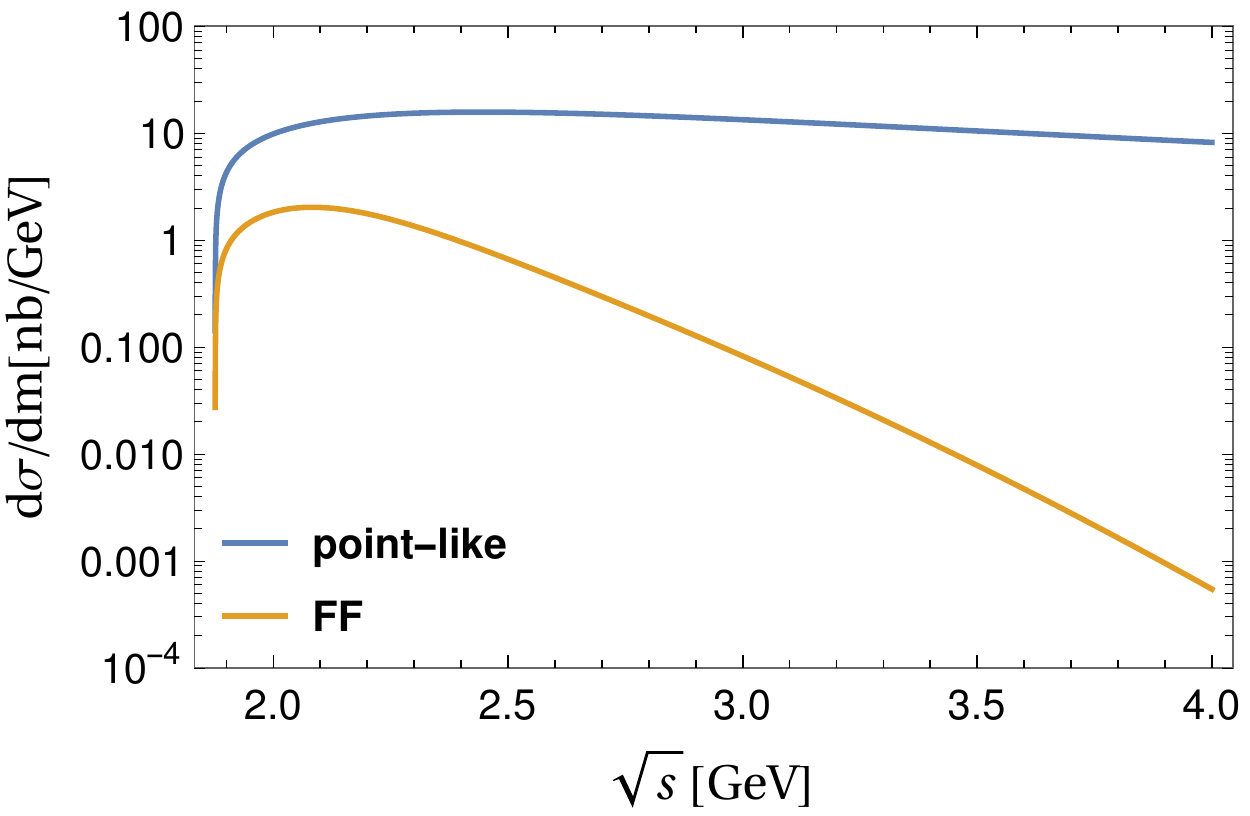}
\caption{\emph{Left:} $|G_{eff}|^2$ in double proton production via  $e^+e^-$ scattering,  $e^+e^-\to p\bar p$, comparing data with the phenomenological model described in the text. \emph{Right:} Contribution of proton exchange mechanism to the cross section of $\gamma\gamma\to p\bar{p}$  for $|\cos\theta|<0.6$. }
\label{fig:proton}
\end{figure}

This model describes quite well data, both in the space-like region ($e^-p\to p e^+$) as well as in the time-like region ($e^+e^-\to p\bar{p}$), of one photon exchange.
$|G_{eff}|^2$ obtained from these expressions is shown on the left panel of \fig{fig:proton}. The parameters used for the fit are extracted from Ref.~\cite{Iachello:2004aq,Bijker:2004yu}. A large enhancement in the kinematical forbidden region can be noticed, this region is not interesting for the proton case but is accessible in a partial compositeness scenario.
Many other sophisticated calculations have addressed the estimate of the nucleon form factors, for example~\cite{Hippchen:1991rr,Mergell:1995bf,Belushkin:2006qa,Alarcon:2017lhg,Bauer:2012pv}.

%%%%%%%%%%%%%%%%%%%%%%%%%%%%%%%%%

For the case of proton pair production via photon scattering we use the model of Ref.~\cite{Klusek-Gawenda:2017lgt} to describe $\gamma\gamma\to p\bar{p}$ from ultraperipheral ion collision.
The model is based on three ingredients: 1- proton exchange; 2- resonances exchange; and 3- handbag mechanism. 
We here are interested only in the proton  exchange mechanism to serve as a basis for our model of gluon interaction.

The computation of the proton exchange is done in two steps. First, the 2 Feynman diagrams ($t$ and $u$ exchange) are summed with the usual $\gamma-p$ interaction given by the Dirac and Pauli form factors \eq{emff}, $\mathcal{M}_{bare}$. To account for the offshellness of the proton propagation, a common form factor is multiplied by the amplitude along the expressions in \eq{eq:ggampff}.
This common form factor guarantees gauge invariance and crossing symmetry and has been successfully used  in several previous works.
The parameter $\Lambda_p$ is fitted to experimental data and is $\Lambda\sim 1\GeV\gtrsim m_p$. The resulting amplitudes fit very well to data once summed to the resonant and handbag contributions and is highly suppressed compared to the point-like prediction. For illustrative purpose we show on the right panel of \fig{fig:proton} the differential cross section resulting from the proton exchange amplitude integrated in the range $|\cos\theta|<0.6$ with the parameters that fit data ($\Lambda_p=1.1$, $\kappa_p=1.7928$) compared to the point-like case.

\bibliographystyle{JHEP}
\bibliography{bib}
\end{document}